\documentclass{article}

\usepackage{arxiv}

\usepackage[utf8]{inputenc} 
\usepackage[T1]{fontenc}    
\usepackage{hyperref}       
\usepackage{url}            
\usepackage{booktabs}       
\usepackage{amsfonts}       
\usepackage{nicefrac}       
\usepackage{microtype}      
\usepackage{lipsum}		
\usepackage{graphicx}
\usepackage[numbers]{natbib}
\usepackage{doi}
\usepackage{enumitem}
\setlist{leftmargin=3.0mm}

\title{Parametrization of Non-Bonded Force Field Terms for Metal–Organic Frameworks Using Machine Learning Approach}

\author{ \href{https://orcid.org/0000-0001-6117-5662}{\includegraphics[scale=0.07]{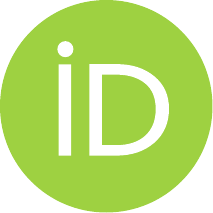}\hspace{1mm}Vadim V. Korolev}\thanks{\textit{Email address}: \texttt{korolev@colloid.chem.msu.ru}} \\
	Department of Chemistry\\
	Lomonosov Moscow State University\\
	Moscow 119991, Russia\\
	\And
	\href{https://orcid.org/0000-0001-7008-4586}{\includegraphics[scale=0.07]{orcid.pdf}\hspace{1mm}Yurii M. Nevolin} \\
	Frumkin Institute of Physical Chemistry\\
	and Electrochemistry\\
	Russian Academy of Sciences\\
	Moscow 119071, Russia\\
	\And
	\href{https://orcid.org/0000-0002-4033-9864}{\includegraphics[scale=0.07]{orcid.pdf}\hspace{1mm}Thomas A. Manz} \\
	Department of Chemical \& Materials Engineering\\
	New Mexico State University\\
	Las Cruces, New Mexico 88003-8001, United States\\
	\And
	\href{https://orcid.org/0000-0002-1503-3679}{\includegraphics[scale=0.07]{orcid.pdf}\hspace{1mm}Pavel V. Protsenko} \\
	Department of Chemistry\\
	Lomonosov Moscow State University\\
	Moscow 119991, Russia\\
}

\renewcommand{\shorttitle}{Parametrization of Non-Bonded Force Field Terms}

\hypersetup{
pdftitle={Parametrization of Non-Bonded Force Field Terms for Metal–Organic Frameworks Using Machine Learning Approach},
pdfsubject={q-bio.NC, q-bio.QM},
pdfauthor={Vadim V.~Korolev, Yurii M.~Nevolin, Pavel V.~Protsenko, and Vladimir G.~Sergeyev},
pdfkeywords={First keyword, Second keyword, More},
}

\begin{document}
\maketitle

\begin{abstract}
The enormous structural and chemical diversity of metal–organic frameworks (MOFs) forces researchers to actively use simulation techniques as often as experiments. MOFs are widely known for their outstanding adsorption properties, so a precise description of the host–guest interactions is essential for high-throughput screening aimed at ranking the most promising candidates. However, highly accurate \textit{ab initio} calculations cannot be routinely applied to model thousands of structures due to the demanding computational costs. Furthermore, methods based on force field (FF) parametrization suffer from low transferability. To resolve this accuracy–efficiency dilemma, we applied a machine learning (ML) approach. The trained models reproduced the atom-in-material quantities, including partial charges, polarizabilities, dispersion coefficients, quantum Drude oscillator and electron cloud parameters, within the accuracy of the underlying density functional theory method. The aforementioned FF precursors make it possible to thoroughly describe non-covalent interactions typical for MOF–adsorbate systems: electrostatic, dispersion, polarization, and short-range repulsion. The presented approach can also readily facilitate hybrid atomistic simulations/ML workflows.
\end{abstract}


\section{Introduction}
Metal–organic frameworks (MOFs) are soft solids that form the most extensive subspace of the nanoporous materials genome.\cite{boyd2017computational} Their building blocks—metal ions/clusters and organic linkers—are assembled into edge-transitive nets, governed by reticular chemistry rules.\cite{yaghi2003reticular,yaghi2019introduction,jiang2021reticular} The structural variety of MOFs gives rise to diverse physical behavior.\cite{mezenov2019metal} Some structures possess unconventional properties for soft matter, including high electrical conductivity,\cite{ko2018conductive,xie2020electrically} superconductivity,\cite{zhang2017theoretical,huang2018superconductivity} and exotic topological band structures.\cite{jiang2021exotic} However, the keen interest in MOFs is mainly due to the outstanding adsorption properties. In particular, their ultrahigh porosity enables record-breaking volumetric and gravimetric uptakes,\cite{chen2020balancing} whereas specific adsorption sites help to capture the target molecule selectively.\cite{boyd2019data} MOFs appear promising for the storage and separation of a wide range of gases and their mixtures, including hydrogen,\cite{suh2012hydrogen} methane,\cite{he2014methane,mason2014evaluating} carbon dioxide,\cite{sumida2012carbon,yu2017co2} and noble gases.\cite{banerjee2015potential,banerjee2018xenon} Unfortunately, complete experimental characterization of a representative candidate set is technically infeasible,\cite{ongari2020too} since the number of synthesized MOFs has reached one hundred thousand to date.\cite{moghadam2017development} The hypothetical structures generated \textit{in silico} are even more numerous.\cite{wilmer2012large,aghaji2016quantitative} For this reason, computational studies have been carried out to reveal the structure–property relationships in MOFs as often as experiments.

The accuracy–efficiency dilemma is especially acute for MOFs due to their hybrid organic–inorganic nature and the relatively large sizes of the unit cells (the typical number of atoms is hundreds or even thousands). The level of theory used to describe host–guest interactions depends on the specific task faced by researchers; a broad set of approximations, differing in electronic coarse graining, have been applied.\cite{odoh2015quantum,mancuso2020electronic} \textit{Ab initio} methods based on Møller–Plesset second-order perturbation theory (MP2) and coupled cluster (CC) approaches provide accurate binding energies.\cite{grajciar2010water,dzubak2012ab,yu2013combined,howe2017acid,barnes2019multilayer} Due to the high computational demands, MOF–adsorbate systems are represented as cluster models that contain adsorption sites and gas molecules, resulting in a loss of a reliable description of the dispersion interactions. The hybrid quantum mechanics/molecular mechanics (QM/MM) approach has been successfully adopted to solve this issue.\cite{yu2013combined} Density functional theory (DFT), a workhorse of computational materials science, has been intensively used to model the adsorption properties of MOFs as well. However, most of the popular exchange-correlation (XC) functionals based on the generalized gradient approximation (GGA) do not account for intermolecular interactions properly. Therefore, long-range dispersion correction plays a critical role in the modeling of MOFs within the DFT framework. There are several generations of the empirical scheme proposed by Grimme and coworkers, which are usually labeled as D1,\cite{grimme2004accurate} D2,\cite{grimme2006semiempirical} D3,\cite{grimme2010consistent} and D4.\cite{caldeweyher2017extension} The van der Waals density functional (vdW-DF) method\cite{berland2015van} implemented in the growing set of XC functionals\cite{larsen2017libvdwxc} captures the vdW forces \textit{via} a nonlocal correlation component. General trends of the adsorption of carbon dioxide,\cite{rana2012comparing,poloni2012ligand,hou2013understanding,queen2014comprehensive,poloni2014understanding,tan2015competitive,lee2015small,mann2016first,vlaisavljevich2017performance,asgari2018experimental} methane,\cite{poloni2014understanding,tan2015competitive,lee2015small,vlaisavljevich2017performance} water,\cite{tan2015competitive,vlaisavljevich2017performance,you2018tuning} and noble gases\cite{vazhappilly2016computational,kancharlapalli2019confinement} in a series of isostructural MOFs have been revealed by employing DFT-based studies.

\textit{Ab initio} and DFT methods cannot provide the scalability required for screening large MOF subsets. A few exceptions are related to the calculations of the intrinsic properties of structures,\cite{chung2014computation,nazarian2016comprehensive,rosen2021machine} regardless of their interactions with adsorbates. Thus, classical simulation techniques, such as the grand canonical Monte Carlo (GCMC) method, provide a theoretical basis for the high-throughput screening (HTS) of small-molecule adsorption in MOFs.\cite{boyd2019data,wilmer2012large,aghaji2016quantitative,simon2015materials,simon2015best,banerjee2016metal,thornton2017materials,ahmed2019exceptional} In these studies, host–guest interactions are described \textit{via} the non-bonded terms of force fields (FFs),\cite{dubbeldam2019design} i.e., interaction potentials. The Universal Force Field (UFF)\cite{rappe1992uff} and DREIDING\cite{mayo1990dreiding} are the most popular generic FFs in MOF studies, but they have several well-known drawbacks. In particular, the polarization effects of the adsorbate molecules induced by open metal sites present a significant challenge for conventional FFs.\cite{pham2014capturing,pham2017predictive,becker2018potential} Extended versions of the UFF\cite{addicoat2014extension,coupry2016extension} and more specialized FFs\cite{bureekaew2013mof,bristow2014transferable} have also been proposed. \textit{Ab initio} derived\cite{vanduyfhuys2015quickFF,haldoupis2015ab,jawahery2019ab} and explicitly polarizable\cite{pham2015understanding,becker2017polarizable,becker2018polarizable} FFs help to significantly improve the description of intermolecular interactions only for a small series of isoreticular structures, leaving the aforementioned dilemma largely unaddressed. Therefore, to facilitate HTS adsorption studies, it is necessary to develop a fast automatized procedure for the generation of FF components that will be suitable for the various atomic types present in MOFs. In the rigid framework approximation, only non-bonded FF terms are needed.

Recently, Chen and Manz\cite{chen2019collection} have presented a collection of non-bonded FF precursors that can be implemented to fully describe non-covalent interactions that are typical for MOF–adsorbate systems: electrostatic, dispersion, polarization, and short-range repulsion. Within this framework, partial charges calculated by the Density Derived Electrostatic and Chemical (DDEC)\cite{manz2016introducing,limas2016introducing,manz2017introducing,limas2018introducing} approach are used to define Coulombic interactions. Dispersion in the dipole approximation is described \textit{via} fluctuating polarizabilities and dispersion coefficients ${C}_{6}$.\cite{manz2019new} Non-directionally screened polarizabilities are intended to incorporate interactions between induced dipoles and external electric fields, charged atoms, permanent multipoles, or other induced dipoles into polarizable FFs.\cite{manz2019new} In the quantum Drude oscillator (QDO) parametrization scheme,\cite{jones2013quantum,sadhukhan2016quantum,cipcigan2019electronic} (many-body) multipole dispersion and polarization interactions are set through the corresponding QDO parameters: mass, frequency, and charge. The electron cloud parameters\cite{van2016beyond} are applicable to describe short-range exchange repulsion.

Several recent studies\cite{korolev2020transferable,zou2020efficient,raza2020message,kancharlapalli2021fast} have partially achieved the fast generation of the FF components using a computational approach beyond atomistic simulations. Thus, machine learning (ML) algorithms make it possible to predict partial charges in MOFs within the accuracy of the underlying DDEC approach. At the same time, ML techniques are comparable for empirical charge equilibration (Qeq)\cite{ongari2018evaluating} methods in terms of scalability.

In this study, we applied a data-driven approach to derive a full suite of atom-in-material quantities required for advanced FF parametrization. Taking a collection of high-quality FF precursors extracted for 3056 MOFs as initial data, we implemented gradient boosting models on top of a diverse set of features that described the local site environment. The combination of a state-of-the-art approximation algorithm and a data representation scheme outperformed previous approaches for partial charge assignment. The trained models for other FF precursors demonstrated high accuracy in terms of the correlation coefficients. The relative contributions of these features to the model performance were estimated by means of two methods, including a game-theoretic approach. In addition, we outlined future opportunities for the presented ML approach and its alternative practical applications beyond FF parametrization.

\section{Materials and Methods}
\label{sec:Methods}

\subsection{Reference Database}
We used a collection of 3056 MOFs\cite{chen2019collection} as a starting dataset. Each structure included the atomic coordinates and the corresponding FF precursors. For further consideration, nine FF precursors were selected: atomic partial charge, dispersion coefficient ${C}_{6}$, and fluctuating and non-directionally screened (FF) polarizabilities. Three parameters described the QDO model (reduced mass, effective frequency, and pseudoelectron effective charge), and there were two electron cloud parameters. Manz and Chen\cite{chen2019collection} extracted partial charges and electron cloud parameters using the DDEC6 partitioning scheme.\cite{manz2016introducing,limas2016introducing,manz2017introducing,limas2018introducing} The dispersion coefficients, polarizabilities, and QDO parameters were calculated using the MCLF method.\cite{manz2019new,manz2019new2} The selected FF precursors represent the minimum set required to describe all non-bonded interaction terms thoroughly.

\subsection{Featurization Scheme}
The properties of an atomic site, beginning with forces\cite{feynman1939forces} and including atom-in-material parameters, are functions of its local environment. In this study, the following diverse set of chemical and structural features was used as input data for the approximation algorithm (ML model):

\begin{itemize}
\item The set of descriptors inspired by the electronegativity equalization principle was originally implemented by Kancharlapalli et al.\cite{kancharlapalli2021fast} We used its extended version (referred to as ENFingerprint) that included the electronegativity and first ionization energy of the considered atomic site, averaged electronegativity of the sites in the first and second coordination sphere, averaged first ionization energy, distances between the target atomic site and sites in its first and second coordination sphere, and the corresponding numbers of sites. The first and second coordination shell included sites that formed a bond with the considered site directly and through one of its nearest neighbors, respectively. Two sites were considered to be bonded if the interatomic distance did not exceed the sum of the corresponding Cordero covalent radii,\cite{cordero2008covalent} within a penalty distance of 0.5 Å. The thermochemical scale\cite{tantardini2021thermochemical} of the dimensionless electronegativities was used.
\item The adaptive generalizable neighborhood informed (AGNI)\cite{botu2015learning,botu2015adaptive} fingerprints are integrals of the product of the radial distribution function and the damping function \({f}_{d}\):

\begin{equation}
  V_{i}(\eta)=\sum _{i \neq j} e^{-(r _{ij}/ \eta)^2} f_{d} (r_{ij})
\end{equation}

\begin{equation}
  f_{d} (r_{ij})=0.5
  \left[ \frac{cos(\pi r_{ij})}{R_{c}}+1 \right]
\end{equation}

where ${r}_{ij}$ is the distance between sites ${i}$ and ${j}$, ${R}_{c}$ is the cutoff distance, and ${\eta}$ is the Gaussian function width.

\item Voronoi-tessellation-based\cite{okabe2016spatial,peng2011structural,wang2019transferable} fingerprints summarize the features of the Voronoi cells, including the Voronoi indices, the (weighted) ${i}$-fold symmetry indices, and the Voronoi volume, area, and nearest-neighbor distance statistics (mean, standard deviation, and minimum and maximum values).

\item CrystalNNFingerprint\cite{zimmermann2017assessing,zimmermann2020local} and OPSiteFingerprint\cite{zimmermann2017assessing,zimmermann2020local} were described as site environment \textit{via} the coordination likelihoods and specific local structure order parameters (LoStOPs). CrystalNN\cite{zimmermann2020local} and the minimum distance\cite{zimmermann2017assessing} neighbor-finding algorithms were used, respectively.

\end{itemize}

All the aforementioned fingerprints were concatenated into a 109-dimensional vector (the full list of fingerprints is provided in the Supporting Information). The crystal structure processing routines were carried out using the Python Materials Genomics (pymatgen)\cite{ong2013python} and Atomic Simulation Environment (ASE)\cite{larsen2017atomic} modules. AGNI, Voronoi, CrystalNN, and OPSite fingerprints were calculated by the matminer\cite{ward2018matminer} library.

Some structures failed to assign one of two (or both) subsets of features during the featurization: Voronoi and ENFingerprint. The issue related to the Voronoi tessellation could be resolved by increasing the cutoff radius when determining the Voronoi neighbors (a default value of 6.5 Å was applied). However, ENFingerprint could not be used for structures containing small (with the longest path in molecular graph no more than three) ions, such as H\textsubscript{3}O\textsuperscript{+}, NH\textsubscript{3}\textsuperscript{+}, and NO\textsubscript{3}\textsuperscript{--}. This is because there was no second coordination sphere for non-central atoms in such ions. Thus, charged MOFs were naturally excluded from further analysis as well. Finally, unique sites from 2946 structures were used to train/validate ML models.

\subsection{Machine Learning (ML) Algorithm}
Within a local approximation, FF precursors are defined by the site’s fingerprints. These generally unknown functions are approximated by the refined implementation of the gradient boosting algorithm, eXtreme Gradient Boosting (XGBoost).\cite{chen2016xgboost} The Classification and Regression Tree (CART) model, which is a tree ensemble model ${\phi}$ as a superposition of ${K}$ additive functions ${f}$ represents the true value of a target property ${y}$ for the ${i}$-th site in the following form:

\begin{equation}
  \widehat{y}_{i}=\phi (x _{i})=\sum _{k=1}^{K}f_{k}(x_{i})
\end{equation}

where ${x}$ is the site’s representation. The parameters of the CART model (tree structure and leaf weights) are fitted during the iterative minimization of regularized objective $\mathcal{L}$:

\begin{equation}
  \mathcal{L}(\phi)=\sum _{i}l(\widehat{y}_{i},{y}_{i})+\sum _{k}\Omega(f_{k})
\end{equation}

where ${l}$ is the differentiable loss function, and ${\Omega}$ is the penalty term designed to avoid over-fitting \textit{via} regularization of the model’s weights.
The XGBoost models were trained to predict each FF precursor independently. Preliminarily, the calculated fingerprints were scaled using MinMaxScaler and were rounded jointly with target values to the fourth decimal; duplicated data were excluded from consideration. We tested models on an external test set (10\% of points from the initial set) with the five-fold cross-validation. The optimal values of the hyperparameters (including the number of gradient boosted trees, maximum tree depth, and boosting learning rate) were determined using the Tree-structured Parzen Estimator (TPE)\cite{bergstra2011algorithms} algorithm implemented in the Hyperopt\cite{bergstra2013hyperopt,bergstra2015hyperopt} library.

\section{Results and Discussion}
\label{sec:ResultsAndDiscussion}

\subsection{Performance of ML Models}
The parity plots of the calculation results of Chen and Manz\cite{chen2019collection} and the ML-predicted FF precursors are reported in Figure 1. The corresponding histograms of the deviations of the predicted values from the reference values are presented in Figure S1. Table 1 summarizes the performances of the trained ML models intended for FF precursor prediction. The most commonly used regression metrics—mean absolute error (MAE), root mean square error (RMSE), and coefficient of determination (${R}^{2}$)—are shown here. The Pearson and Spearman coefficients, which measure the linear and rank correlation, respectively, are also provided. In general, a high density of points near the diagonal of the parity plots and high values (> 0.96) of the three coefficients (${R}^{2}$, Pearson, and Spearman) indicate the superior predictive capabilities of the presented models. However, these metrics do not provide insights into efficiency for specific modeling tasks \textit{per se}. In other words, it is unclear whether the presented models simulate the adsorption properties of MOFs with sufficient accuracy.

\begin{table}[ht]
\centering
\caption{\label{tab:Table1}Summary of performances of presented machine learning (ML) models.}
\begin{tabular}{ |p{6cm}||c|c|c|c|c|}
 \hline
 FF precursor & MAE & RMSE & ${R}^{2}$ & Pearson & Spearman\\
 \hline
 partial charge ${q}$ & 0.0113 & 0.0216 & 0.9970 & 0.9985 & 0.9960\\
 fluctuating polarizability ${log}_{10}({\alpha}_{fluc})$ & 0.0095 & 0.0159 & 0.9977 & 0.9989 & 0.9905\\
 FF polarizability ${log}_{10}({\alpha}_{FF})$ & 0.0070 & 0.0126 & 0.9982 &   0.9991 & 0.9917\\
 dispersion coefficient ${log}_{10}({C}_{6})$ & 0.0134 & 0.0217 & 0.9990 & 0.9995 & 0.9923\\
 QDO mass ${m}_{QDO}$ & 0.0090 & 0.0196 & 0.9976 & 0.9988 & 0.9918\\
 QDO charge ${q}_{QDO}$ & 0.0042 & 0.0067 & 0.9985 & 0.9993 & 0.9928\\
 QDO frequency ${\omega}_{QDO}$ & 0.0081 & 0.0129 & 0.9794 & 0.9897 & 0.9863\\
 electron cloud parameter ${a}$ & 0.0554 & 0.0871 & 0.9816 & 0.9908 & 0.9828\\
 electron cloud parameter ${b}$ & 0.0225 & 0.0358 & 0.9627 & 0.9814 & 0.9785\\
 \hline
\end{tabular}
\end{table}

\begin{figure}[ht]
  \centering
  \includegraphics[height=13cm]{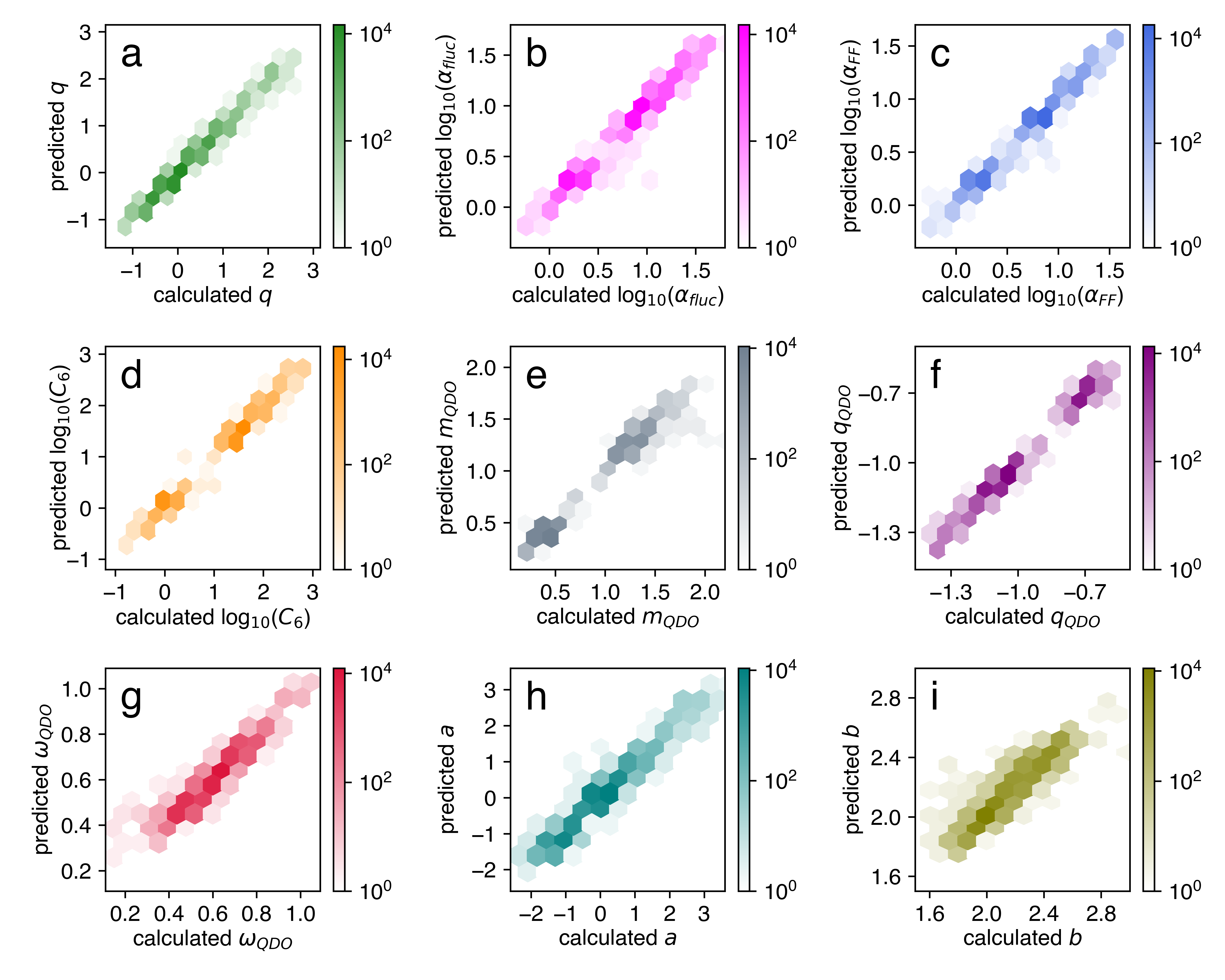}
  \caption{Parity plots of the calculated results of Chen and Manz\cite{chen2019collection} and ML-predicted FF precursors: (a) partial charge, (b) fluctuating polarizability, (c) FF polarizability, (d) dispersion coefficient ${C}_{6}$, (e) QDO mass, (f) QDO charge, (g) QDO frequency, (h) electron cloud parameter ${a}$, and (i) electron cloud parameter ${b}$.}
  \label{fig:fig1}
\end{figure}

This difficulty can be partially resolved by comparing our models with those available in the literature. The following ML approaches were used to predict the partial charges in MOFs: the multilayer connectivity-based atom contribution (m-CBAC) method developed by Zou et al.,\cite{zou2020efficient} the message passing neural networks (MPNNs) implemented by Raza et al.,\cite{raza2020message} random forest models in conjunction with features inspired by the electronegativity equalization principle (PACMOF\cite{kancharlapalli2021fast} package), and our implementation\cite{korolev2020transferable} that included local structural features as inputs to the XGBoost models. The direct comparison of the listed approaches is hindered by the differences in the used partitioning scheme (DDEC3\cite{manz2012improved} versus DDEC6\cite{manz2016introducing,limas2016introducing,manz2017introducing,limas2018introducing}) and the sets of MOFs. In addition, although in all these studies the number of structures under consideration was about three thousand, the data sizes differed significantly. It is well known that the availability of materials data has a significant impact on the predictive precision of ML models.\cite{zhang2018strategy} Therefore, the following estimates are general and are not true performance benchmarks. In terms of the MAE, the presented partial charge predictor, with an MAE of 0.0113 e\textsuperscript{--}, slightly outperformed the PACMOF\cite{kancharlapalli2021fast} and MPNN\cite{raza2020message}, with MAEs of 0.0192 and 0.025 e\textsuperscript{--}, respectively. The less representative Pearson and Spearman coefficients are given for the m-CBAC\cite{zou2020efficient} approach. Their values (0.997 and 0.984, respectively) were lower than those obtained in this study (0.9985 and 0.9960). The only close competitor was our previous implementation,\cite{korolev2020transferable} which showed an even smaller MAE of 0.0096 e\textsuperscript{--}. The insignificant difference may have been due to the distinction in the featurization schemas and, more importantly, the removal of duplicate data in this study.

The aforementioned approaches\cite{korolev2020transferable,zou2020efficient,raza2020message,kancharlapalli2021fast} have been validated by comparing values of the adsorption properties calculated using ML-derived and DDEC charges. Thus, the Spearman rank coefficient between the CO\textsubscript{2} Henry coefficients computed using DDEC and ML-derived charges (obtained by the m-CBAC\cite{zou2020efficient} approach and MPNNs\cite{raza2020message}) equaled 0.939 and 0.96, respectively. The Spearman rank coefficient for the CO\textsubscript{2} volumetric uptakes computed using the DDEC and ML-derived charges presented in our previous study\cite{korolev2020transferable} was 0.991. PACMOF\cite{kancharlapalli2021fast} could reproduce the CO\textsubscript{2} loading, N\textsubscript{2} loading, and CO\textsubscript{2}/N\textsubscript{2} selectivity with mean absolute percentage errors (MAPEs) of 18.9, 28.3, and 33.9, respectively. Thus, the models that yielded MAEs of 0.01–0.02 e\textsuperscript{--} were sufficiently accurate to reproduce the values of the adsorption properties obtained using the DDEC charges. In this context, the Spearman rank correlation coefficient is much more representative than in the case of partial charge prediction, since ranking promising candidates for a specific application can be seen as a primary goal of HTS studies. Similar estimates for other FF precursors are not available.

Another aspect of the ML model’s efficiency concerns how its accuracy relates to the reference method. In computational chemistry, the so-called chemical accuracy ($\sim$1 kcal/mole) usually serves as a desirable level of precision for reproducing potential energy surfaces (PESs) by \textit{ab initio} methods. Recently, the same can be said about ML models trained on calculated data.\cite{bogojeski2020quantum,kalita2021learning} From a more general perspective, the following guiding principle can be formulated: the accuracy of an approximation model that relies on quantum-chemical inputs should be at least comparable to the accuracy of the underlying computational method relative to the experimentally measured quantities. As for FF precursors, extracting experimental values is quite complex and not straightforward, so such an analysis can be carried out for a very limited set of available data. Thus, the MAE deviations of the DDEC6 charges from those of charges derived \textit{via} kappa refinement\cite{hansen1978testing,coppens1979net,zuo2004measurements} for natrolite and formamide were 0.1174 and 0.0570 e\textsuperscript{--}, respectively.\cite{limas2016introducing} The MCLF method yielded the static polarizability tensor eigenvalues for six small organic molecules and dispersion coefficients ${C}_{6}$ for pairs formed from 49 atoms/molecules within mean absolute relative errors (MAREs) of 8.10\% and 4.45\%, respectively.\cite{manz2019new2} The static polarizabilities and dispersion coefficients ${C}_{6}$ for 12 polyacenes defined by this method had MAREs within 8.75\% and 7.77\%, respectively. The values for six fullerenes were 5.92\% and 6.84\%, respectively.\cite{manz2019new2} Our XGBoost models reproduced the   fluctuating polarizabilities extracted using the MCLF method and the reference dispersion coefficients ${C}_{6}$with the following MAREs: 2.18\% and 3.07\%, respectively. We speculate the accuracies of the presented approximation algorithms are comparable to the precision of the reference methods, DDEC6 and MCLF.

\subsection{ML Model Interpretability}
The selection of a reliable material representation is an essential step in constructing a precise predictive model.\cite{ghiringhelli2015big,ramprasad2017machine,butler2018machine} The initial choice of descriptors is usually based on domain expertise. For instance, previous studies that aimed at partial charge assignment using ML techniques used a small set of physically meaningful parameters\cite{kancharlapalli2021fast} or a collection of atomic connectivity-based patterns.\cite{zou2020efficient} In this study, we used a different approach, applying a diverse suite of heterogeneous fingerprints. The validity of this approach, i.e., non-redundancy of chosen features, can be confirmed by conducting feature importance analysis. As a result, the revealed input parameters that do not contribute to the model’s output can be reasonably excluded from consideration. First, we calculated analog of one of the most popular feature importance measures of ensemble learners, also known as Gini importance (“gain” in XGBoost implementation).\cite{breiman1984classification} This quantity is defined as the mean decrease in impurity (MDI), which is the sum of all decreases in the Gini impurity over all trees in the ensemble. The normalized values of the gain-based importance for five feature subsets are shown in Figures 2a, 2c, and 2e. ENFingerprint made a decisive contribution: its importance varied from 83.5\% to 95.9\% for the electron cloud parameters ${a}$ and ${b}$, respectively. Based on these data, it can be concluded that all other subsets had a negligible impact on the performances of the ML models. To confirm this, we re-trained models to predict the partial charges using only the ENFingerprint as an input (importance of 95.2\%). Surprisingly, the MAE increased from 0.0113 (see Table 1) to 0.0185 e\textsuperscript{--}, i.e., it grew by 63\%. Such a dramatic decrease in the model accuracy was inconsistent with the minor importance of the four other fingerprint subsets determined by a gain-based method and was likely due to its intrinsic shortcomings, including sampling bias.\cite{strobl2007bias}

\begin{figure}[ht]
  \centering
  \includegraphics[height=7.5cm]{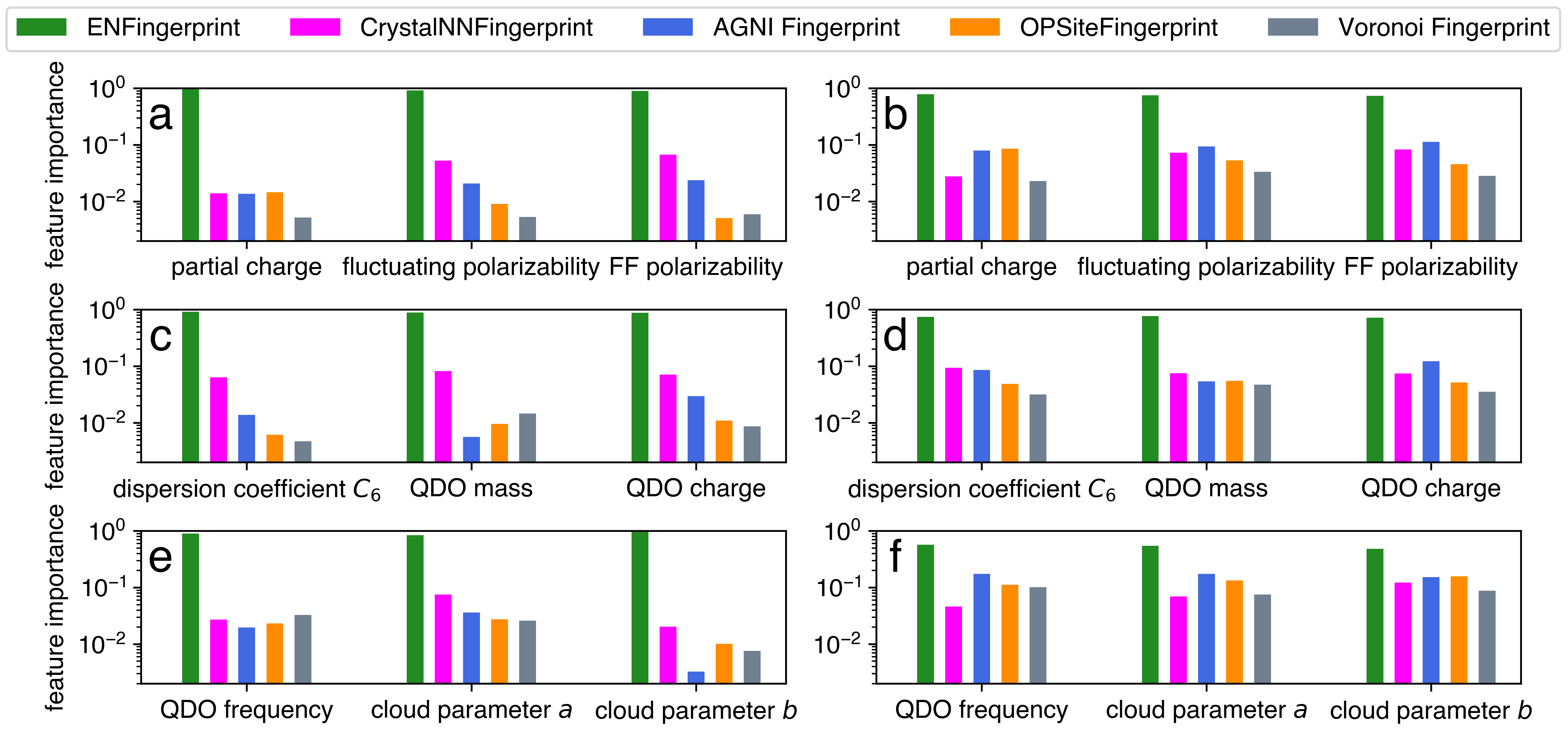}
  \caption{Cumulative feature importances corresponding to fingerprint subsets. The reported values are obtained by (a, c, and e) gain-based method and (b, d, and f) SHAP analysis.}
  \label{fig:fig2}
\end{figure}

We then calculated the SHAP (SHapley Additive exPlanations) values using the TreeExplainer\cite{lundberg2020local} algorithm to obtain a more reliable estimate of the feature importance. These quantities represent an extended version of the classical Shapley values,\cite{shapley1953value} which originate from game theory. Explanations of the predictions expressed using the SHAP values are guaranteed to satisfy the following properties: local accuracy, monotonicity, and missingness. Global feature attribution was carried out by averaging the magnitudes of the SHAP values over all testing set points since TreeExplainer provides local object-wise explanations. Mean SHAP values normalized to unity over all features are provided in Figures 2b, 2d, and 2f. The impact of ENFingerprint was significantly decreased compared to gain-based feature importance values and varied from 48.0\% (electron cloud parameter ${b}$) to 78.6\% (partial charge). The cumulative importance for each of the other fingerprint subsets reached 10\% for at least one FF precursor. Thus, the used suite of features should be considered non-redundant and reasonable.

\subsection{Future Opportunities}
In addition to feature representation and approximation algorithm, training data availability also determines the accuracy of ML predictors.\cite{zhang2018strategy} We trained a series of models on datasets that differed in size (from one to three hundred thousand atomic sites) to extract this dependency for three low-correlated targets: partial charge, fluctuating polarizability, and QDO frequency. The corresponding dataset size vs. scaled error dependencies are presented in Figure 3. The power law $SE=m\times{DS}^{-n}$ almost perfectly describes all three sets of points, where ${SE}$ is the scaled error (MAE normalized by the range of the corresponding FF precursor), ${DS}$ is the training data size, and ${m}$ and ${n}$ are empirical parameters. It should be noted that coefficient ${n}$ (the slope of a line in logarithmic coordinates) differs in each case: 0.27 ± 0.03, 0.20 ± 0.03, and 0.17 ± 0.04 for the partial charge, fluctuating polarizability, and QDO frequency, respectively. The given values are significantly lower than those obtained for a diverse set of properties\cite{zhang2018strategy} (0.372) and the formation energy of perovskites\cite{schmidt2017predicting} (0.297). Therefore, the universal dependency derived by Zhang and Ling\cite{zhang2018strategy} was not observed, at least for the atom-in-material quantities considered in this study. Nevertheless, the revealed power law suggests that the FF precursors’ accuracy can be improved extensively by increasing the training data size.

\begin{figure}[b]
  \centering
  \includegraphics[height=6.5cm]{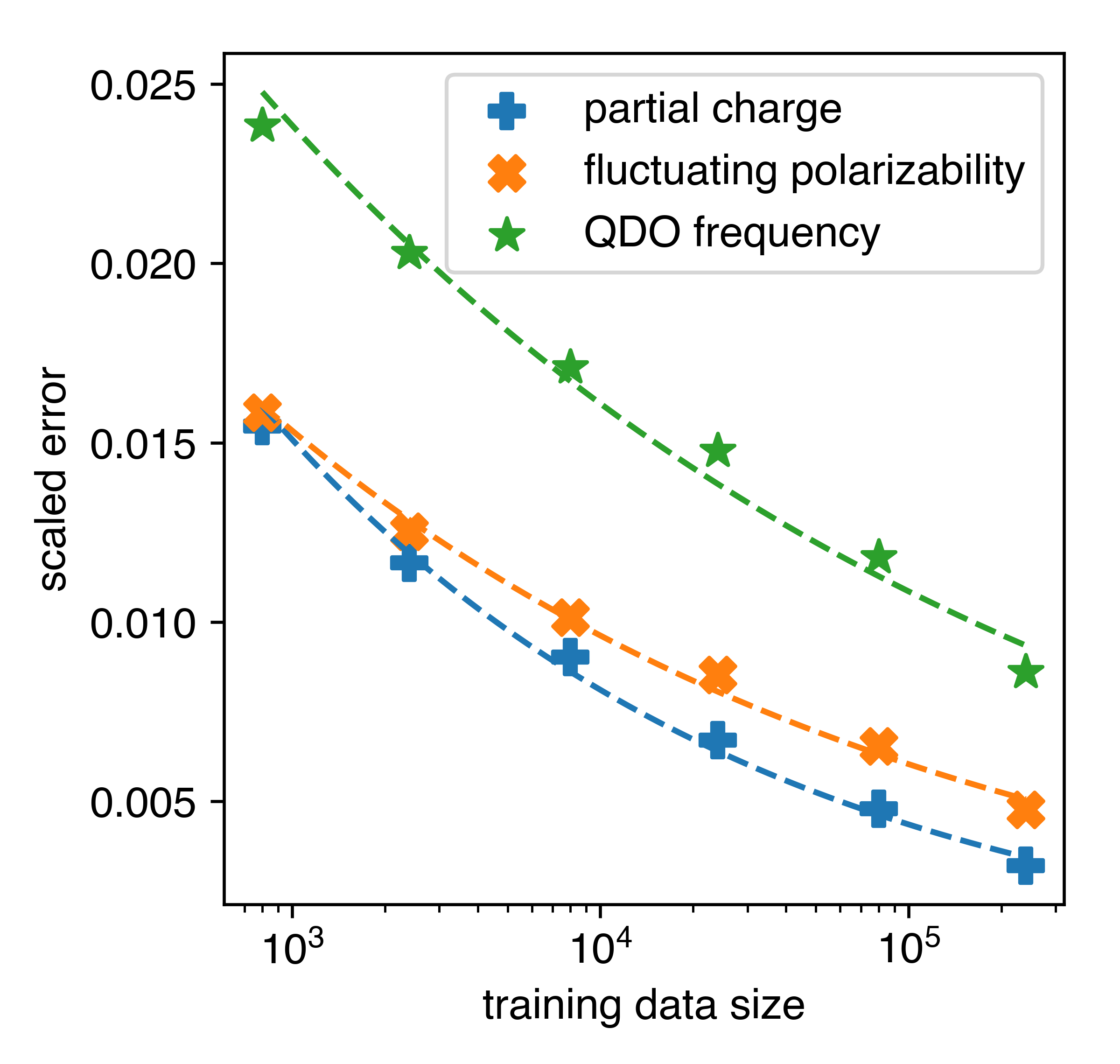}
  \caption{Scaled error as a function of training data size.}
  \label{fig:fig3}
\end{figure}

As indicated in the Methods section, ML models for each FF precursor were trained independently. However, due to the use of a common representation of the atomic site, multi-task learning (MTL)\cite{caruana1997multitask} framework can be efficiently applied here. The performance of the primary method for MTL, deep neural networks (DNNs), improves in the presence of highly correlated targets.\cite{xu2017demystifying} To assess the potential of MTL for FF precursor predictions, we calculated Pearson coefficients for all pairs of the considered atom-in-material quantities. The correlation matrix in the form of a heatmap is shown in Figure 4. Two groups of highly correlated FF precursors can be distinguished. The first group includes the fluctuating polarizability, FF polarizability, dispersion coefficient ${C}_{6}$, and electron cloud parameter a. The second group contains the QDO mass, QDO charge, and electron cloud parameter b. Therefore, MTL predictors for the listed endpoints can potentially outperform the corresponding single-task models.

The presented models can also be helpful to facilitate the ML prediction of adsorption properties. Data-driven approximations are at best able to reproduce the quantitative structure–property relationships hidden in the input data but still inheriting errors specific to the underlying computational approach.  Thus, most ML models (as opposed to atom-wise predictors) that were aimed at predicting macroscopic adsorption properties\cite{altintas2021machine,jablonka2020big} were trained on results of GCMC simulations, for which UFF is almost no alternative choice for describing non-covalent interactions. Therefore, the outputs of those ML models suffered from the issues mentioned in the Introduction section, such as lacking a description of the polarization effects. The following hybrid workflow can improve the reliability of the predicted targets without losing scalability: advanced parametrization using a full suite of FF precursors (main scope of this study) → HTS adsorption modeling in rigid framework approximation → construction of ML predictors of macroscopic properties.

\begin{figure}[t]
  \centering
  \includegraphics[height=8.0cm]{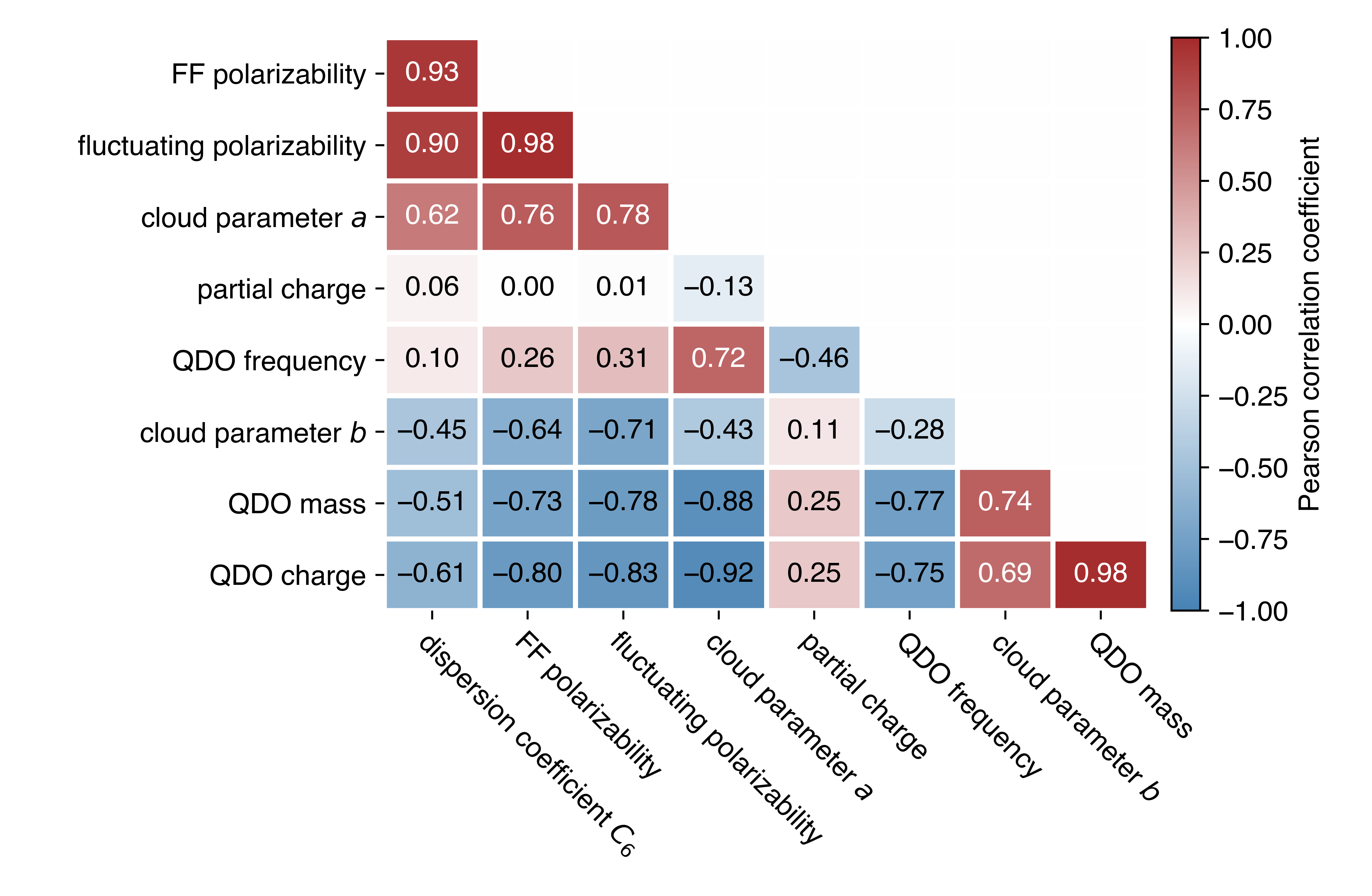}
  \caption{Correlation matrix for considered FF precursors.}
  \label{fig:fig4}
\end{figure}

\section{Conclusions}
In summary, we presented the ML workflow to reproduce atom-in-material quantities useful for the parametrization of FFs. Its modular structure, typical for data-driven approaches, rests on three pillars: input data, feature representation, and approximation algorithm. Each of the parts can be modified depending on the specific task. In principle, our approach is also applicable for other subclasses of nanoporous materials, including covalent organic frameworks (COFs) and hydrogen-bonded organic frameworks (HOFs). Since the transferability of the presented models to structures beyond MOFs is in question, reliable results can be obtained using reference DDEC and MCLF data derived for a specific subclass of materials under consideration. The set of local features, i.e., the input for the ML algorithm, is modifiable as well. However, it is highly desirable to confirm the validity of the new set based on feature importance, as was demonstrated in this study. Finally, a reasonable choice of the approximation algorithm requires full-fledged benchmarking that takes into account accuracy and time efficiency.

\section{Data and Software Availability}
All MOF crystal structures and corresponding FF precursors used to train XGBoost models are available as electronic supplementary information (ESI) at \url{https://doi.org/10.1039/C9RA07327B}. The full pipeline, including featurization and FF precursor prediction, is shared through GitHub as an open-source python library, FFP4MOF (\url{https://github.com/korolewadim/ffp4mof}). The trained XGBoost models are available on Zenodo using the DOI \url{https://doi.org/10.5281/zenodo.5500641}.

\section{Acknowledgments}
The authors thank Prof. Vladimir G. Sergeyev (Lomonosov Moscow State University) for helpful comments. This study was funded by the Ministry of Science and Higher Education of Russian Federation, project number 121031300084-1.

\bibliographystyle{unsrt}
\bibliography{references}

\end{document}